\documentclass[12pt]{article}
\usepackage{amsmath}
\usepackage{mathrsfs}
\usepackage{graphicx}
\usepackage{ amssymb }
\usepackage{multirow}
\usepackage[margin=2cm]{geometry} 
\usepackage{slashed}
\usepackage{caption}

\def\un#1{\relax\ifmmode\@@underline#1\else
        $\@@underline{\hbox{#1}}$\relax\fi}

\let\du=\du                     

\def\a{\alpha}
\def\b{\beta}
\def\c{\chi}

\def\g{\gamma}

\def\m{\mu}
\def\n{\nu}

\def\p{\pi}

\def\ve{\varepsilon}

\def\bo{{\raise-.3ex\hbox{\large$\Box$}}}               
\def\pa{\partial}                                       
\def\TH{{\raise.2ex\hbox{$\displaystyle \bigodot$}\mskip-4.7mu \llap H \;}}
\def\face{{\raise.2ex\hbox{$\displaystyle \bigodot$}\mskip-2.2mu \llap {$\ddot
        \smile$}}}                                      

\def\leftrightarrowfill{$\mathsurround=0pt \mathord\leftarrow \mkern-6mu
        \cleaders\hbox{$\mkern-2mu \mathord- \mkern-2mu$}\hfill
        \mkern-6mu \mathord\rightarrow$}
\def\dvec#1{\vbox{\ialign{##\crcr
        \leftrightarrowfill\crcr\noalign{\kern-1pt\nointerlineskip}
        $\hfil\displaystyle{#1}\hfil$\crcr}}}           

\newskip\humongous \humongous=0pt plus 1000pt minus 1000pt
\def\caja{\mathsurround=0pt}
\def\eqalign#1{\,\vcenter{\openup2\jot \caja
        \ialign{\strut \hfil$\displaystyle{##}$&$
        \displaystyle{{}##}$\hfil\crcr#1\crcr}}\,}
\newif\ifdtup

 \newcommand{\be}{\begin{equation}}
\newcommand{\ee}{\end{equation}}
\newcommand{\nbe}{\begin{equation*}}
\newcommand{\nee}{\end{equation*}}

\newcommand{\lb}{\label}

\def\fracmm#1#2{{{#1}\over{#2}}}
  
\def\low#1{{\raise -3pt\hbox{${\hskip 0.75pt}\!_{#1}$}}}

\begin{document}

\thispagestyle{empty}

{\hbox to\hsize{
\vbox{\noindent July 2018 \hfill IPMU18-0123 }}
\noindent  \hfill }

\noindent
\vskip2.0cm
\begin{center}

{\Large\bf Beyond Starobinsky inflation
}

\vglue.3in

Yermek Aldabergenov~${}^{a,b}$, Ryotaro Ishikawa~${}^{a}$, Sergei V. Ketov~${}^{a,b,c}$ and
Sergey I. Kruglov~${}^{d}$
\vglue.1in

${}^a$~Department of Physics, Tokyo Metropolitan University, \\
Minami-ohsawa 1-1, Hachioji-shi, Tokyo 192-0397, Japan \\
${}^b$~Research School of High-Energy Physics, Tomsk Polytechnic University,\\
2a Lenin Ave., Tomsk 634050, Russian Federation \\
${}^c$~Kavli Institute for the Physics and Mathematics of the Universe (IPMU),
\\The University of Tokyo, Chiba 277-8568, Japan \\
${}^d$~Department of Chemical and Physical Sciences, University of Toronto, 3359 Mississauga
Road, North, Mississauga, Ontario, Canada L5L 1C6 
\vglue.1in

aldabergenov-yermek@ed.tmu.ac.jp, ishikawa-ryotaro@ed.tmu.ac.jp, ketov@tmu.ac.jp,
serguei.krouglov@utoronto.ca
\end{center}

\vglue.3in

\begin{center}
{\Large\bf Abstract}
\end{center}
\vglue.1in
\noindent  A supergravity extension of the $(R+R^2)$ gravity with the additional (Born-Infeld) 
structure of a massive vector multiplet gives rise to the specific $F(R)$ gravity, whose 
structure is investigated in detail. The massive vector multiplet has inflaton (scalaron), goldstino and
a massive vector field as its field components.  The model describes Starobinsky inflation and allows us to extrapolate the $F(R)$ function beyond the inflationary scale (up to Planck scale). We observe
some differences versus the $(R+R^2)$ gravity, and several breaking patterns of the well known correspondence between the $F(R)$ gravity and the scalar-tensor gravity.

\newpage

\section{Introduction}

An ultimate theory of cosmological inflation should be based on quantum gravity and is yet to be constructed. This is related to another open problem of finding an {\it ultraviolet} (UV) completion of any phenomenologically viable inflationary model. Amongst the most successful and popular inflationary models, Starobinsky inflationary model of $(R+R^2)$ gravity \cite{Starobinsky:1980te} is special because it is entirely based on gravitational interactions. This model is, however, non-renormalizable and has the UV-cutoff given by Planck scale. In addition, when extrapolating the $(R+R^2)$ gravity beyond the inflationary scale of about $10^{13}$ GeV, i.e. when going to the very large curvature regime, we are left with the scale-invariant $R^2$ gravity. The original motivation  in \cite{Starobinsky:1980te} was to get rid of the initial singularity of Einstein-Friedmann gravity, in addition to describing inflation in the early Universe. However, demanding the asymptotical scale invariance at very high energies is clearly not the only option. Hence, is still the open question: what should we expect beyond Starobinsky inflation? 

To address this question at least partially, one needs a motivated extension of the $(R+R^2)$ gravity in a specific framework. In this paper, we address the issue in four-dimensional $N=1$ supergravity. The importance of the inflationary model building in supergravity stems from the natural objective to unify gravity with particle physics beyond the Standard Model of elementary particles and beyond the Standard ($\Lambda$CDM) Model of cosmology, see e.g., \cite{Yamaguchi:2011kg,Ketov:2012yz} for a review.

Though supergravity can be considered as the low-energy effective theory of (compactified) superstrings, and the latter can be viewed as a consistent theory of quantum gravity, we obviously need more specific assumptions.

Our additional specific assumptions in this paper are the following:
\begin{itemize}
\item Starobinsky inflationary model  should be embedded into a four-dimensional $N=1$ supergravity, with linearly realized (manifest) local supersymmetry,
\item inflaton (scalaron) should belong to a massive $N=1$ vector supermultiplet,
\item the kinetic terms of the vector supermultiplet should have the Born-Infeld (or Dirac-Born-Infeld) structure, inspired by superstrings and D-branes.
\end{itemize}

This leads to the specific (modified) $F(R)$ gravity model, whose peculiar structure is in the focus of our investigation in this paper.

Our paper is organized as follows. In Sec. 2 we outline Born-Infeld (BI) non-linear electrodynamics and the supergravity theory with the BI structure. In Sec. 3 we review the Starobinsky inflationary model. In Sec. 4 we study in detail the  $F(R)$ gravity extension of the $(R+R^2)$ gravity, originating from the  supergravity theory. In Sec.~5 we present the dual description of the same $F(R)$ gravity in terms of the scalar-tensor gravity. Our Conclusion is Sec.~6. In Appendix we formulate the full supergravity theory in terms of superfields in curved superspace.

\section{Born-Infeld structure in gravity and supergravity}

Born-Infeld (BI) Lagrangian was originally introduced \cite{Born:1934gh} as a non-linear generalization of the Lagrangian of Maxwell electrodynamics in terms of the abelian field strength 
$F_{\m\n}=\pa_{\m}A_{\n} -\pa_{\n}A_{\m}$,
\be \lb{bi} 
L_{\rm BI} = -b^{-2}\left[ \sqrt{-\det\left(\eta_{\m\n}+\frac{b}{e}F_{\m\n}\right)} -1\,\right]=
-\fracmm{1}{4e^2}F^{\m\n}F_{\m\n} + {\cal O}(F^4)~~,
\ee
where we have introduced the dimensional (BI) coupling constant $b=M^{-2}_{\rm BI}$ and
the gauge (dimensionless) coupling constant $e$. Being {\it minimally} coupled to gravity, the BI action reads
\be \lb{big} 
S_{\rm BI} = b^{-2}\int d^4x\,\left[ \sqrt{-g} - \sqrt{-\det\left(g_{\m\n}+\frac{b}{e}F_{\m\n}\right)} \,\right]~.
\ee

This BI structure also arises (i) in the bosonic part of the open superstring effective action \cite{Fradkin:1985qd}, (ii) as part of Dirac-Born-Infeld effective action of a D3-brane \cite{Leigh:1989jq}, and (iii) as part of Maxwell-Goldstone action describing partial supersymmetry breaking of $N=2$ supersymmetry to $N=1$ supersymmetry \cite{Bagger:1996wp,Rocek:1997hi}. In string theory, $b=2\pi\alpha'$, while the BI scale $M_{\rm BI}$ does not have to coincide with $M_{\rm Pl}$.~\footnote{See also  \cite{Ketov:2001dq,Ketov:1996bm} for more about special properties of the BI action and its supersymmetric extensions.}

In $N=1$ supersymmetry and supergravity, a vector field belongs to an $N=1$ vector multiplet, whose
supergravity couplings are naturally (off-shell) described in superconformal tensor calculus \cite{Freedman:2012zz} and in curved superspace \cite{Wess:1992cp}. A {\it massive} $N=1$ vector multiplet has a single (real) scalar field amongst its bosonic field components, in addition to a massive vector field. In this paper, we identify this real scalar with inflaton, and unify it with the massive vector field whose kinetic terms are assumed to have the BI structure in $N=1$ supergravity (we
do not assume any relation between our massive vector field and electromagnetic field). 

The full action of the self-interacting massive vector multiplet with the BI structure in supergravity is very complicated: it was found by using the superconformal tensor calculus in 
\cite{Abe:2015fha}, and we present this action in Appendix, by using superfields in curved superspace.~\footnote{See also \cite{Aldabergenov:2016dcu,Aldabergenov:2017bjt,Addazi:2017ulg} for related papers.}  In particular, local supersymmetry (SUSY) is spontaneously broken in this theory (after inflation also), while goldstino is identified with a massive "photino" in the same vector multiplet with inflaton. 
 
 For our purposes in this paper, it is enough to notice that in the dual (modified supergravity) picture the BI structure just leads to the presence of the contribution $12R^2/(e^2M^4_{\rm BI})$  under the square root of the BI term, in addition to the $F_{\m\n}$-dependent terms there. When ignoring all other interactions besides the modified gravity itself (i.e. keeping only the $R$-dependent terms),  it gives rise to the following  $F(R)$ gravity model (see Ref.~\cite{Abe:2015fha} and Appendix):
\be \lb{mgr}
S = \int d^4x \sqrt{-g}\left[ \fracmm{M^2_{\rm Pl}}{2} R + \fracmm{M^4_{\rm BI}}{3} \left(
 \sqrt{ 1 + \fracmm{12R^2}{e^2M^4_{\rm BI}} } -1\right)\,\right]~~. 
\ee

It is this modified gravity theory that is the main subject of our investigation in this paper. It is worth
noticing that it does {\it not\/} imply the upper bound on the values of $R$, unlike the original BI theory 
(\ref{bi}) that limits the maximal values of the gauge field strength components.

It is worth noticing here that the idea of finding a "BI-extension" of Einstein gravity is old but still popular, although it lacks a good definition and guiding principles, see e.g., \cite{BeltranJimenez:2017doy} for classification of many such extensions in gravitational theory, and \cite{Kruglov:2014gva} for other proposals to an $F(R)$ gravity function
of the BI-type.

A "BI-extension" of $N=1$ supergravity is more restrictive, but it suffers similar problems, see e.g., \cite{Gates:2001ff} for some specific proposals of BI supergravity in curved superspace. Equation (\ref{mgr}) is just the specific extension of Starobinsky $(R+R^2)$ gravity in the framework of $F(R)$ gravity derived  from supergravity and inspired by string theory. It is directly related to the BI action (\ref{bi}) that arises together with the $F(R)$ gravity (\ref{mgr}) in the same supergravity theory having the BI structure.

It is also worth mentioning that Starobinsky inflation is equivalent to the so-called Higgs inflation in gravity and supergravity, because both lead to the same inflationary observables \cite{Ketov:2012jt}.

\section{Starobinsky inflation and $F(R)$ gravity}

Starobinsky model of inflation is defined by the action \cite{Starobinsky:1980te}
 \be \label{star}
S_{\rm Star.} = \fracmm{M^2_{\rm Pl}}{2}\int \mathrm{d}^4x\sqrt{-g} \left( R +\fracmm{1}{6m^2}R^2\right)~,
\ee
where we have introduced the reduced Planck mass $M_{\rm Pl}=1/\sqrt{8\p G_{\rm N}}\approx
2.4\times 10^{18}$ GeV, and the scalaron (inflaton) mass $m$ as the only parameter. We use the
spacetime signature $(-,+,+,+,)$. The $(R+R^2)$ gravity model (\ref{star}) can be considered as the simplest extension of the standard Einstein-Hilbert action in the context of (modified) $F(R)$ gravity theories with an action
 \be \label{fg}
S_F = \fracmm{M^2_{\rm Pl}}{2}\int \mathrm{d}^4x\sqrt{-g} \, F(R)~,
\ee
in terms of the function $F(R)$ of the scalar curvature $R$.

The $F(R)$ gravity action (\ref{fg}) is classically equivalent to 
\begin{equation} \lb{eq}
 S[g_{\m\n},\chi] = \fracmm{M^2_{\rm Pl}}{2} \int d^{4}x \sqrt{- g}~ \left[ F'(\chi) (R - \chi) + F(\chi) \right]
\end{equation}
with the real scalar field $\chi$, provided that $F''\neq 0$ that we always assume.  Here the
primes denote the derivatives with respect to the argument. The equivalence is easy to verify because the $\chi$-field equation implies $\chi=R$. In turn, the factor $F'$ in front of the $R$ in (\ref{eq}) can be (generically) eliminated by a Weyl transformation of metric $g_{\m\n}$, that transforms the action (\ref{eq}) into the action of the scalar field   $\chi$ minimally coupled to Einstein gravity and having the scalar potential 
\be \lb{spot}
 V =      \left(\fracmm{M^2_{\rm Pl}}{2}\right)       \fracmm{\chi F'(\chi) - F(\chi)}{F'(\chi)^{2}}~~.
\ee
Differentiating this scalar potential yields
\be \lb{diffp}
\frac{dV}{d\chi} =  \left(\fracmm{M^2_{\rm Pl}}{2}\right)    
\fracmm{F''(\chi)\left[2F(\chi) - \chi F'(\chi)\right]}{ (F'(\chi))^{3}}~~.
\ee

The kinetic term of $\chi$ becomes canonically normalized after the field redefinition $\chi(\varphi)$ as
\begin{equation} \lb{fred}
  F'(\chi) =  \exp \left(  \sqrt{\frac{2}{3}} \varphi/M_{\rm Pl} \right)~,\quad
  \varphi =  \fracmm{\sqrt{3}M_{\rm Pl}}{\sqrt{2}}\ln F'(\c) ~~,
\end{equation}
in terms of the canonical inflaton field $\varphi$, with the total acton
\be \lb{quint}
S_{\rm quintessence}[g_{\m\n},\varphi]  = \fracmm{M^2_{\rm Pl}}{2}\int \mathrm{d}^4x\sqrt{-g} R 
 - \int \mathrm{d}^4x \sqrt{-g} \left[ \frac{1}{2}g^{\m\n}\pa_{\m}\varphi\pa_{\n}\varphi
 + V(\varphi)\right]~.
\ee

The classical and quantum stability conditions of $F(R)$ gravity theory are given by 
\cite{Ketov:2012yz}
\begin{equation} \lb{stab}
 F'(R) > 0 \quad {\rm and} \quad F''(R) > 0~,
\end{equation}
and they are obviously satisfied for Starobinsky model (\ref{star}) for $R>0$.

Differentiating the scalar potential $V$ in Eq.~(\ref{spot}) with respect to $\varphi$ yields 
\begin{equation} \lb{diff1}
    \frac{d V}{d \varphi}   = \frac{d V}{d \chi} \frac{d \chi}{d \varphi} 
      = \frac{M^2_{\rm Pl}}{2} \left[ \frac{\chi F'' + F' - F'}{F'^{2}} - 2 \frac{\chi F' - F}{F'^{3}} F'' \right] 
\frac{d \chi}{d \varphi}~~,
\end{equation}
where we have
\begin{equation} \lb{diff2}
  \frac{d \chi}{d \varphi}
    = \frac{d \chi}{d F'} \frac{ d F'}{d \varphi}
    = \frac{d F'}{d \varphi} \left/ \frac{d F'}{d\chi} \right.
    = \fracmm{\sqrt{2}}{\sqrt{3}M_{\rm Pl}}  \frac{F'}{F''}~~.
\end{equation}
This implies
\begin{equation} \lb{derv}
  \frac{d V}{d \varphi} = M_{\rm Pl}\frac{2F-\chi F'}{\sqrt{6}  F'^{2}}~~.
\end{equation}
Combining Eqs.~(\ref{spot}) and (\ref{derv})  yields $R$ and $F$ in terms of the scalar potential $V$,
\begin{align} \lb{inv}
  & R = \left[  \fracmm{\sqrt{6}}{M_{\rm Pl}}
    \frac{d V}{d \varphi} + \fracmm{4V}{M^2_{\rm Pl}} \right] \exp \left(  \sqrt{\frac{2}{3}} 
  \varphi/M_{\rm Pl} \right),  \\
  & F= \left[  \fracmm{\sqrt{6}}{M_{\rm Pl}}
 \frac{d V}{d \varphi} + \fracmm{2V}{M^2_{\rm Pl}} \right] \exp \left(  2 \sqrt{\frac{2}{3}} \varphi/M_{\rm Pl}\right).
\end{align}
These equations define the function $F(R)$ in the parametric form, in terms of a scalar potential $V(\varphi)$, i.e. the {\it inverse} transformation to (\ref{spot}). This is known as the classical equivalence (duality) between the $F(R)$ gravity theories (\ref{fg}) and the scalar-tensor (quintessence) theories of gravity (\ref{quint}).

In the case of Starobinsky model (\ref{star}), one gets the famous potential
\begin{equation} \label{starp}
V(\varphi) = \fracmm{3}{4} M^2_{\rm Pl}m^2\left[ 1- \exp\left(-\sqrt{\frac{2}{3}}\varphi/M_{\rm Pl}\right)\right]^2~.
\end{equation}
This scalar potential is bounded from below (non-negative and stable), and it has the absolute  minimum at 
$\varphi=0$  corresponding to  a Minkowski vacuum. The scalar potential (\ref{starp}) also has a {\it plateau} of positive height (related to inflationary energy density), that gives rise to slow roll of inflaton in the inflationary era.
 The Starobinsky model (\ref{star}) is the particular case of the so-called $\alpha$-attractor inflationary models \cite{Galante:2014ifa}, and is also a member of the close family of viable inflationary models of $F(R)$ gravity, originating from higher dimensions \cite{Nakada:2017uka}.

A duration of inflation is measured in the slow roll approximation by the e-foldings number
\be \lb{efold}
N_e\approx  \fracmm{1}{M^2_{\rm Pl}} \int_{\varphi_{\rm end}}^{\varphi_{*}} \fracmm{V}{V'}d\varphi~~,
\ee
where $\varphi_{*}$ is the inflaton value at the reference scale (horizon crossing), and $\varphi_{\rm end}$ is the
inflaton value at the end of inflation when one of the slow roll parameters
\be \lb{slowp}
\ve_V(\varphi) = \fracmm{M^2_{\rm Pl}}{2}\left( \fracmm{V'}{V}\right)^2 \quad {\rm and} \quad 
\eta_V(\varphi) = M^2_{\rm Pl} \left( \fracmm{V''}{V}\right)~~,
\ee
is no longer small (close to 1).

The amplitude of scalar perturbations at horizon crossing is given by \cite{Ellis:2015pla}
\be \lb{amp}
A = \fracmm{V_*^3}{12\p^2 M^6_{\rm Pl}({V_*}')^2}=\fracmm{3m^2}{8\p^2M^2_{\rm Pl}}\sinh^4\left(
\fracmm{\varphi_*}{\sqrt{6}M_{\rm Pl}}\right)~~.
\ee

The Starobinsky model  (\ref{star}) is the excellent model of cosmological inflation, in very good agreement with the Planck data \cite{Ade:2015xua,Ade:2015lrj,Array:2015xqh}. The Planck satellite mission measurements of the Cosmic Microwave Background (CMB) radiation \cite{Ade:2015xua,Ade:2015lrj,Array:2015xqh}
give the scalar perturbations tilt as $n_s\approx 1+2\eta_V -6\ve_V\approx 0.968\pm 0.006$ and restrict the 
tensor-to-scalar ratio as $r\approx 16\ve_V < 0.08$. The Starobinsky inflation yields $r\approx 12/N_e^2\approx 0.004$ and $n_s\approx 1- 2/N_e$, where $N_e$ is the e-foldings number between 50 and 60, with the best fit at $N_e\approx 55$ \cite{Mukhanov:1981xt,Kaneda:2010ut}.

The Starobinsky model (\ref{star}) is geometrical (based on gravity only), while its (mass) parameter $m$ is fixed by the 
observed CMB amplitude (COBE, WMAP) as 
\be \lb{starm}
m\approx 3 \cdot10^{13}~{\rm GeV} \quad {\rm or}\quad \fracmm{m}{M_{\rm Pl}}\approx 1.3\cdot 10^{-5}~.
\ee
A numerical analysis of (\ref{efold}) with the potential (\ref{starp}) yields \cite{Ellis:2015pla}
\be\lb{sandf}
\sqrt{\fracmm{2}{3}} \varphi_*/M_{\rm Pl} \approx \ln\left( \fracmm{4}{3}N_e\right) \approx 5.5\quad {\rm and}
\quad \sqrt{\fracmm{2}{3}} \varphi_{\rm end}/M_{\rm Pl} \approx \ln\left[ \fracmm{2}{11}(4+3\sqrt{3})\right]\approx 0.5~~,
\ee
where we have used $N_e\approx 55$.

\section{BI-modified Starobinsky model}

In accordance to (\ref{fg}), the modified gravity theory (\ref{mgr}) has 
\be \lb{mgrf}
F(R) = R  + \fracmm{2g^2}{3\b}\left( \sqrt{1+12\b R^2} -1\right)~~,
\ee
where we have introduced the parameters $g=1/(eM_{\rm Pl})$ and $\b=1/(e^2M^4_{\rm BI})$. In
this parametrization, our $F$-function (\ref{mgrf}) exactly agrees with Eq.~(37) of Ref.~\cite{Abe:2015fha}.

When assuming $12\b R^2\ll 1$, the function (\ref{mgrf}) gives rise to the $(R+R^2)$ gravity model of
Starobinsky in (\ref{star}), as it should. It allows us to identify
\be \lb{eg}
g^2=\fracmm{1}{24m^2}\quad {\rm and} \quad e^2=24\left(\fracmm{m}{M_{\rm Pl}}\right)^2\approx
4\cdot 10^{-9}~~,
\ee
where we have used (\ref{starm}). In terms of the dimensionless quantities
$\tilde{F} =F/M^2_{\rm Pl}$ and $\tilde{R} = R/M^2_{\rm Pl}$, and the dimensionless parameters 
\be \lb{dsp}
\a = \fracmm{M_{\rm BI}}{M_{\rm Pl}}\quad {\rm and} \quad \tilde{\g}=e\a^2~~,
\ee
we have the dimensionless function
\be \lb{dsfg}
\tilde{F}(\tilde{R})= \tilde{R} + \fracmm{2}{3}\a^4  \left( \sqrt{1+12\tilde{R}^2/\tilde{\g}^2} -1\right)
\ee

A global shape of this function is given in Fig.~1.


\begin{figure}[t]
\begin{center}
\vspace{1cm}
\includegraphics[width=15cm,height=10cm]{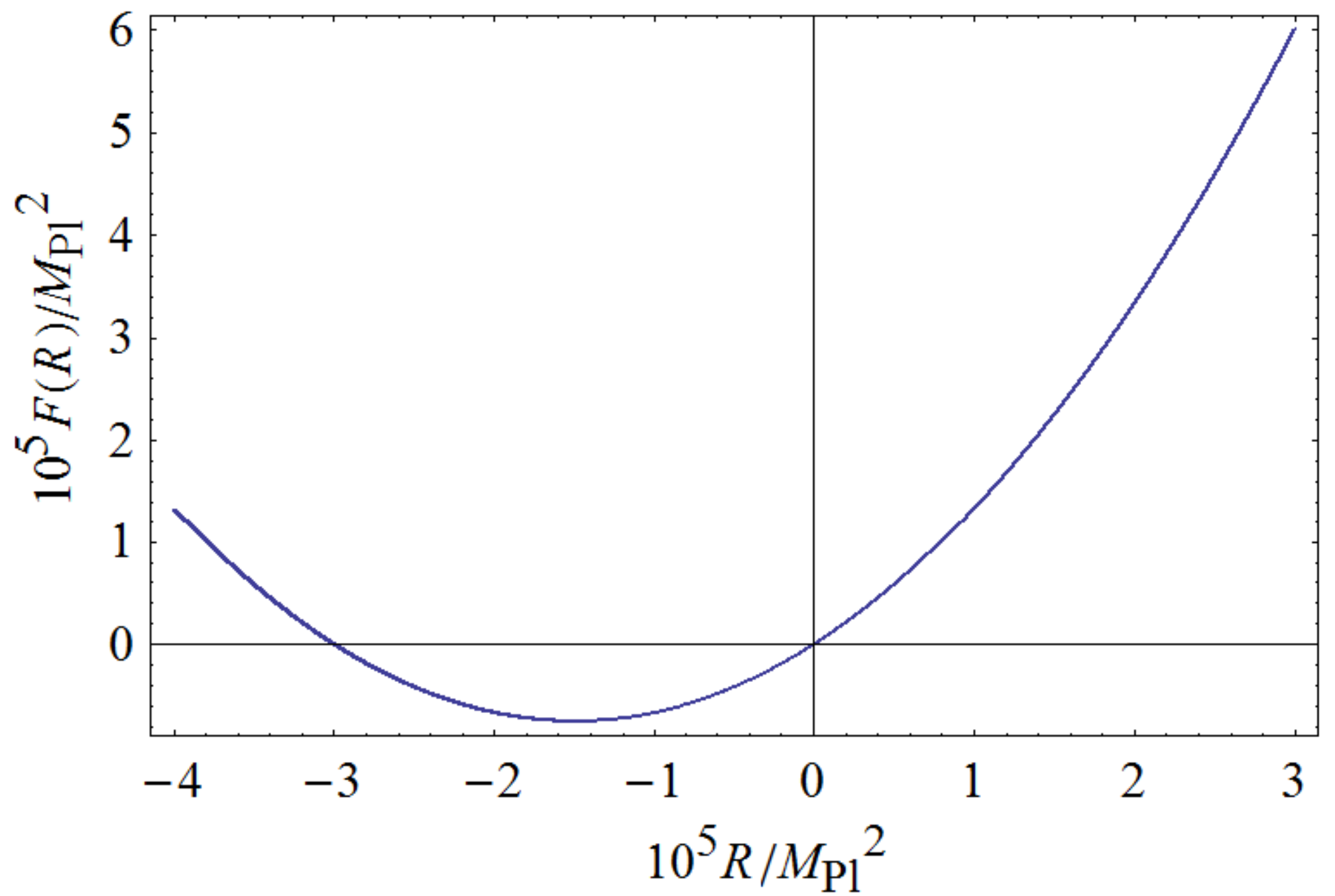}
\caption{the profile of the $F(R)$ gravity function (\ref{mgrf}) for $\a=1$ and 
 $\tilde{\g}^{-2}=10^{5}$. This value of the parameter $\tilde{\g}$ is only chosen to demonstrate the global shape of the function.} \label{fig:1}
\end{center}
\end{figure}

The physical conditions imply the range $\tilde{R}\in [-1,1]$ (i.e. up to the UV-cutoff) and  $\a\in [0.01,1]$ (i.e. between the Grand Unification scale and Planck scale), so that $\tilde{\g}\in 6.3\cdot[10^{-7},10^{-5}]$. The Starobinsky inflation takes place for $0<\tilde{R}\ll \tilde{\g}$.

\newpage

The function (\ref{mgrf}) is well defined for {\it any} values of $R$, and implies three physical regimes:
\begin{itemize}
\item the small curvature regime, where gravity is described by the standard Einstein-Hilbert action,
\item the inflationary regime, where gravity is described by Starobinsky $(R+R^2)$ action (\ref{star}),
\item the high curvature regime, where gravity is again described by the Einstein-Hilbert action,
though with the different (larger) effective Planck scale $M_{\rm Pl, effective}=M_{\rm Pl}
\left(1+4g^2/\sqrt{3\b}\right)^{1/2}\leq 189M_{\rm Pl}$ for large positive values of $R$.
\end{itemize}

Static solutions to the $F(R)$ gravity field equations with $R = const.\equiv R_0$ follow from our
equations (\ref{diffp}) and (\ref{derv}), and are given by solutions to the algebraic equation \cite{Barrow:1983rx}
\be \lb{stat}
RF'(R)=2F(R)~~,
\ee
In our case (\ref{mgrf}), with
\begin{equation}
F'(R)=1+\frac{8g^2R}{\sqrt{1+12\beta R^2}} >0 \quad {\rm for}\quad R\geq 0~,
\label{3}
\end{equation}
we find
\begin{equation}
\frac{8g^2R^2_0}{\sqrt{1+12\beta R_0^2}}=R_0+\frac{4g^2}{3\beta}\left(\sqrt{1+12\beta R_0^2}-1\right)
\label{4}
\end{equation}
that gives rise to the condition
\begin{equation}
R_0\left[4(16g^4-3\beta)R_0^3+32g^2R_0^2-R_0+\frac{8g^2}{3\beta} \right]=0.
\label{5}
\end{equation}

Besides the trivial solution $R_0=0$ corresponding to a stable Minkowski vacuum, any other real positive solution ($R_0>0$) must obey the cubic  equation
\begin{equation}
aR_0^3+bR_0^2+cR_0+d=0,
\label{6}
\end{equation}
whose coefficients are $a=4(16g^4-3\beta)$, $b=32g^2$, $c=-1$ and $d=8g^2/(3\beta)$. By using
the standard replacement
\begin{equation}
y=R_0+\frac{b}{3a}~~,
\label{7}
\end{equation}
we can bring (\ref{6}) to the canonical form
\begin{equation}
y^3+3py+2q=0,
\label{8}
\end{equation}
where we have 
\begin{equation}
2q = \frac{2b^3}{27a^3}-\frac{bc}{3a^2}+\frac{d}{a}=\frac{ 4g^2( 1152g^8 -104g^4\beta +27\beta^2)}{27\beta (16g^4-3\beta)^3}~~,
\label{9}
\end{equation}
and
\begin{equation}
3p=\frac{3ac-b^2}{3a^2}=\frac{9\beta-304g^4}{12(16g^4-3\beta)^2}~.
\label{10}
\end{equation}
The number of real solutions depends upon the sign of the cubic discriminant $D=q^2+p^3$ that in our case reads 
\begin{equation}
D = \frac{(144g^4 +\beta)(32g^4+3\beta)^2}{5184 \beta^2 (16g^4-3\beta)^4}~~.
\label{11}
\end{equation}
Since $D>0$, there is only one real solution. Our numerical studies show that this root $R_0$ 
 is negative (e.g., with $\a=1$ we find $R_0\approx - 10^{-7}M_{\rm Pl}^2$).

The second derivative of the $F(R)$ gravity function (\ref{mgrf}) 
 \begin{equation} \lb{16}
F''(R)=\frac{8g^2}{(1+12\beta R^2)^{3/2}} >0
\end{equation}
can be compared to the laboratory bound of E\"ot-Wash experiment \cite{Kapner:2006si} :
$F''(0)\leq 2\times 10^{-6}$ cm$^2$ or
 \begin{equation}
g< 0.5\times10^{-3}\mbox{cm}^2,
\label{17}
\end{equation}
that is well satisfied because of (\ref{starm}) and (\ref{eg}).

\section{Scalar-tensor gravity and inflaton scalar potential}

It is instructive to study the same gravitational model (\ref{mgr}) in the dual (scalar-tensor gravity) picture
defined by (\ref{spot}), (\ref{fred}) and (\ref{quint}). The classical equivalence (duality) between the $F(R)$ gravity
theories and their scalar-tensor gravity (or quintessence) counterparts is well known, see e.g., \cite{mf}.

Our equation (\ref{fred}) implies
\be \lb{rphi}
\fracmm{\tilde{R}}{\tilde{\g}} =
\fracmm{\frac{1}{2}\tilde{\g}\left(1-e^{-\sqrt{2/3}\tilde{\varphi}}\right)}{ \sqrt{
16\a^2 -3\tilde{\g}^2\left( 1-e^{-\sqrt{2/3}\tilde{\varphi}}\right)^2}}~~~,
\ee
where we have introduced the dimensionless inflaton field $\tilde{\varphi}=\varphi/M_{\rm Pl}$. Actually, (\ref{fred})
determines $R^2$ as the function of $\varphi$, and our sign choice in (\ref{rphi}) comes from demanding
a plateau of the scalar potential at positive values of $R$.

In turn, our equation (\ref{spot}) yields
\be \lb{scalarp}
\tilde{V}= \fracmm{\a^4}{3}\sqrt{ 1+12\tilde{R}^2/\tilde{\g}^2}\,
\fracmm{\sqrt{1+12\tilde{R}^2/\tilde{\g}^2}-1}{\left( 8\a^4\tilde{\g}^{-1}(\tilde{R}/\tilde{\g})+
\sqrt{1+12\tilde{R}^2/\tilde{\g}^2}\right)^2}~~~,
\ee
where we have introduced the dimensionless scalar potential $\tilde{V}=V/M^4_{\rm Pl}$. The scalar potential $\tilde{V}(\tilde{\varphi})$ is obtained via a substitution of (\ref{rphi}) into (\ref{scalarp}), while
the value of the parameter $\tilde{\g}$, according to Sections 3 and 4, is given by 
$\tilde{\g}\approx 6.3\cdot 10^{-5}\a^2$.

A profile of the scalar potential is given in Fig.~2.


\begin{figure}[t]
\begin{center}
\vspace{1cm}
\includegraphics[width=15cm,height=8cm]{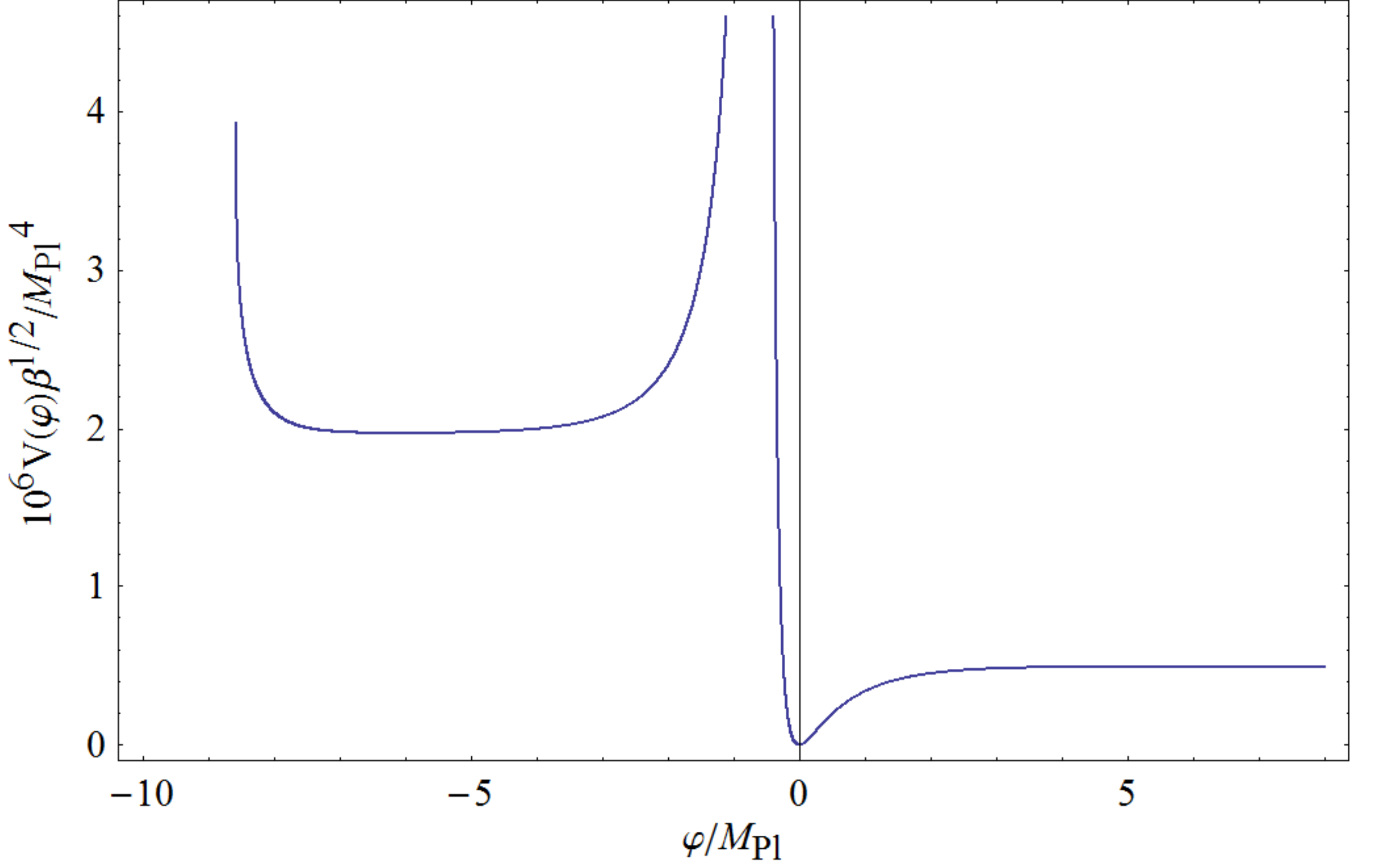}
\caption{the profile of the $V(\varphi)$ function (\ref{scalarp}) for $\a=1$ and $\tilde{\g}=6.3\cdot 
 10^{-5}$. This function is {\it not} well defined for all values of $\tilde{\varphi}$. It reproduces the inflationary potential (\ref{starp}) for the relevant values of $\tilde{\varphi}$ (Sec.~3). The infinite maximum occurs at  $\tilde{\varphi}\approx -0.6$ that corresponds to $\tilde{R}\approx -5\cdot 10^{-10}$. The only anti-de-Sitter minimum occurs at $\tilde{\varphi}\approx -6.5$ that corresponds to the root $\tilde{R}_0\approx -10^{-7}$ found in Sec.~4. The wall on the left-hand-side, where  $V$ sharply goes up to infinity, appears at $\tilde{\varphi}\approx -9$. } \label{fig:2}
\end{center}
\end{figure}

As expected, the scalar potential $V(\varphi)$ has {\it a plateau} for positive values of $\varphi$ and $R$, which corresponds to Starobinsky inflation (Sec.~3). As is clear from (\ref{rphi}), the higher the values of $\varphi$ and $R$ are, the closer the potential $V(\varphi)$ to the Starobinsky potential (\ref{starp}) with $V_{\rm max.}=\fracmm{3}{4}m^2M_{\rm Pl}^2$ is. Hence, the BI structure does not play a significant role for positive values of $\varphi$ and $R$.

When formally sending $\varphi\to +\infty$ in (\ref{rphi}),  we get $\tilde{R}_{\rm max.}=\fracmm{\tilde{\g}^2}{2\sqrt{16\a^2-
3\tilde{\g}^2}}>0$. The scalar-tensor gravity description does {\it not} exist for $\tilde{R}>\tilde{R}_{\rm max.}$, whereas the
$\tilde{F}(\tilde{R})$ gravity description (\ref{dsfg}) is well defined there. This is an explicit example of {\it breaking} 
the naive equivalence between the two dual descriptions.

Though the scalar potential  $V(\varphi)$ {\it cannot} be trusted for large {\it negative} values of
 $\varphi$ and $R$, because of intense particle production (reheating) starting near the absolute minimum of the scalar potential, it is instructive to illustrate two more {\it breaking patterns} of the naive equivalence between $F(R)$ gravity theories and  scalar-tensor gravity theories in our specific example. 

First, we observe the {\it infinite} maximum of the scalar potential in Fig.~2. It happens when the expression under the root in the denominator of (\ref{scalarp}) vanishes,  that corresponds to zero of $F'(R)$ in 
(\ref{spot}) at a negative value of $R$. Since this occurs at a {\it finite} value of $R$, it represents an example of of the broken correspondence, when the $F(R)$ gravity description is regular, but the scalar-tensor description is
singular.

Second, yet another example of the broken correspondence is given by the wall on the left-hand-side of Fig.~2. This wall appears when the expression under the root in the denominator of (\ref{rphi}) vanishes at a {\it finite} value of $\varphi$ that gives rise to the infinite values of $R$ and the scalar potential  $V(\varphi)$, although the value of $V(R)$ remains finite. Beyond the wall, the scalar-tensor gravity description does not exist in our case.
\vglue.2in

\section{Conclusion}

Our main results are given in Sections 4 and 5. They provide a viable extension of Starobinsky
$(R+R^2)$ inflationary model, motivated by the Born-Infeld structure in supergravity, in turn, motivated by string theory. 

Our physical motivation is to explore the range of energies beyond the Starobinsky inflationary
scale of approximately $10^{13}$ GeV up to the (reduced)  Planck scale of approximately 
$10^{18}$ GeV, by using the specific modified gravity function   (\ref{mgr}) derived from the supergravity model under our assumptions formulated in Sec.~1.

The significant deviation between our modified $F(R)$ gravity model and Starobinsky $(R+R^2)$ gravity model takes place only for very large positive curvature, with the asymptotic $R^2$ gravity being replaced by the asymptotic Einstein-Hilbert gravity having a larger effective Planck scale. The corresponding values of the inflaton field are {\it trans-Planckian}, so that the asymptotic gravity  is supposed to be considered with a grain of salt, because it may be affected by quantum gravity effects.

On the other side, we found explicit examples of breaking the naive correspondence between the
$F(R)$ gravity theories and the scalar-tensor gravity theories in our model. They are, however, of academic interest in the inflationary physics context, because they occur at large negative values of the curvature.

\newpage

\section*{Acknowledgements}

YA and SVK are supported in part by the Competitiveness Enhancement Program of Tomsk Polytechnic University in Russia. SVK is also supported in part by a Grant-in-Aid of the Japanese Society for Promotion of Science (JSPS) under No.~26400252,  and the World Premier International Research Center Initiative (WPI Initiative), MEXT, Japan. 
One of the authors (SVK) is grateful to Ignatios Antoniadis for useful discussions.
\vglue.2in

\section*{Appendix: supergravity with BI structure in superspace}

The supersymmetric extension of the $(R+R^2)$ gravity (with Maxwell structure) in the new-minimal formulation of $N=1$ supergravity is given by eq.~(38) of Ref.~\cite{Abe:2015fha} in the superconformal  tensor calculus. In curved superspace, with $M_{\rm Pl}=1$, the Lagrangian reads \cite{Cecotti:1987qe,Farakos:2013cqa} 
\begin{equation}
{\cal L}=\int d^2\Theta 2{\cal E}\left(-\frac{3}{16}\bar{{\cal D}}^2V_{\rm R}+\frac{\gamma}{4}W^\alpha(V_{\rm R})W_\alpha(V_{\rm R})\right)+{\rm h.c.}~,\label{one}
\end{equation}
where $V_{\rm R}$ is the gauge multiplet of SUSY algebra, representing the new-minimal set of supergravity field components, $W_\alpha$ is its superfield strength, and $\gamma\sim e^{-2}$ is the $R^2$ parameter. The superfield $V_{\rm R}$ has the following bosonic components (in a 
Wess-Zumino gauge):
\begin{equation}
\bar{{\cal D}}_{\dot{\alpha}}{\cal D}_\alpha V_{\rm R}\vert=2\sigma_{\dot{\alpha}\alpha}^mA_m~,~~~\bar{{\cal D}}^2{\cal D}^2V_{\rm R}\vert=\frac{32}{3}b_mA^m+16D_{\rm R}~,
\end{equation}
where $A_m$ is the (dynamical) gauge field,
$$D_{\rm R}=\frac{1}{3}\left(R+\frac{3}{2}B_mB^m\right)$$
is the gravitational D-term, and $B_m$ is the auxiliary vector field of supergravity multiplet. 
The old-minimal set of supergravity  is also present via $\cal E$ and $\cal R$ that is hidden in the definition of the superfield strength $W_\alpha\equiv-\frac{1}{4}(\bar{\cal D}^2-8{\cal R}){\cal D}_\alpha V_{\rm R}$. 

After identifying the "old" auxiliary field $b_m$ with the "new" auxiliary field $B_m$ as $b_m=-\frac{3}{2}B_m$, we can
expand the Lagrangian \eqref{one} as follows:
\begin{equation}
e^{-1}{\cal L}=\frac{1}{2}R+\frac{3}{4}B_mB^m-\frac{3}{2}B_mA^m_{\rm}-\frac{1}{4e^2} F_{mn}F^{mn}+\frac{2}{e^2}\left(R+\frac{3}{2}B_mB^m\right)^2+\ldots~,
\end{equation}
where we have kept only the relevant terms. When allowing the superfield $V_{\rm R}$ to be massive (or not using a WZ gauge), the complex scalar $M$ of the old-minimal set \cite{Wess:1992cp} also appears.

The BI extension of the supergravity theory (\ref{one})  can be written down as follows:
\begin{equation} \lb{bione}
{\cal L}=\left(-\frac{3}{16}\int d^2\Theta 2{\cal E}\bar{{\cal D}}^2V_{\rm R}+{\rm h.c.}\right)+\frac{\gamma}{4}\int d^4\theta EW^2\bar{W}^2\Lambda~,
\end{equation}
where the BI structure function $\Lambda$ is given by (see e.g., Ref.~\cite{Ketov:2001dq}) 
\begin{equation}
\Lambda\equiv\frac{\kappa}{1+\kappa(\omega+\bar{\omega})+\sqrt{1+\kappa(\omega+\bar{\omega})+\frac{\kappa^2}{4}(\omega-\bar{\omega})^2}}~,
\end{equation}
with $\omega\equiv{\cal D}^2W^2/8$ and the BI coupling $\kappa=b^{-2}=M^{-4}_{\rm BI}$. The Lagrangian (\ref{bione}) can be expanded as
\begin{equation}
\eqalign{
e^{-1}{\cal L}~~= & ~~~\frac{1}{2}R+\frac{3}{4}B_mB^m-\frac{3}{2}B_mA^m_{\rm}\cr
& +\fracmm{M^{4}_{\rm BI}}{3}
\left(\sqrt{1-\fracmm{3}{2M^4_{\rm BI}e^2}\left(F^2-8(R+\frac{3}{2}B_mB^m)^2\right) + \left(\fracmm{3}{4M^4_{\rm BI}e^2}\right)^2 
(F\tilde{F})^2}-1\right)+\ldots,~\cr}
\end{equation}
where we have kept only the relevant terms. Using $B_m=F_{mn}=0$ as a solution, we get (\ref{mgr}).

\bibliographystyle{utphys} 

\providecommand{\href}[2]{#2}\begingroup\raggedright
\endgroup

\end{document}